\documentclass[11pt]{article}
\usepackage{amsfonts}

\newtheorem{theorem}{Theorem}[section]
\newtheorem{lemma}[theorem]{Lemma}
\newtheorem{corollary}[theorem]{Corollary}

\newcommand{\beginproof}{\medskip\noindent{\bf Proof.~}}
\newcommand{\beginproofof}[1]{\medskip\noindent{\bf Proof of #1.~}}
\newcommand{\finishproof}{\hspace{0.2ex}\rule{1ex}{1ex}}
\newenvironment{proof}{\beginproof}{\unskip\nolinebreak\finishproof\par\medskip }

\newcommand{\state}[2][t]{\mbox{$\phi_{{#2}}^{({#1})}$}}
\newcommand{\pos}[2][\sigma]{\mbox{${{#1}}^{-1}({#2})$}}
\newcommand{\snorm}[1]{\mbox{$|\!|\!|{#1}|\!|\!|_2$}}
\newcommand{\bstr}[1]{\mbox{$\{0, 1\}^{#1}$}}

\long\def\mytitlepage#1#2#3#4{
        \thispagestyle{empty}
        \begin{center}
        {\Large\bf #1}

        \vspace{0.6in}
        #2\\
        \medskip

        \vspace{1in}
        {\Large Abstract}
        \end{center}

        \noindent{#3}
        \vskip60pt

        \noindent{#4}
        \clearpage
        }
\begin{document}
\mytitlepage{
Quantum lower bound for sorting
        \footnote{This work was supported in part by the
                National Science Foundation
                under grant CCR-9820855.}}
{Yaoyun Shi\footnote{
        Department of Computer Science,
        Princeton University,
        Princeton, NJ 08544.
        E-mail: shiyy@cs.Princeton.EDU}
}
{We prove that $\Omega(n\log{n})$
comparisons are necessary for any
quantum algorithm
that sorts $n$ numbers with high success probability and
uses only comparisons. If no error is allowed,
at least $0.110n\log_2{n} - 0.067n + O(1)$ 
comparisons
must be made. The previous known lower bound is $\Omega(n)$.

}
{}

\section{Introduction}
The speedups of quantum algorithms over classical algorithms have 
been the main reason for current interests in quantum computation.
Although dramatic speedups may be possible, for example, Shor's
\cite{Shor97} algorithms for factoring and 
for finding discrete logarithms, the hunt for 
\emph{provable} speedups has been successful only
in very restricted models, such as the decision tree model.

In the decision tree model, the inputs $x_1, x_2, \cdots, x_n$
are known only to an oracle, and the only way that the algorithm
can access the inputs is by asking the oracle questions of the
type ``$x_i = ?$''. The complexity measurement is the number of such
queries.  Many problems that allow \emph{provable} quantum 
speedups can be formulated in this model. For example,
Grover's \cite{Grover96} algorithm for deciding $i$
given the input $e_i$, i.e., the $n$-bits binary string
that has a $1$ in the $i$th position and $0$ elsewhere,
makes $O(\sqrt{n})$
queries, while any classical algorithm needs $\Omega(n)$ queries.
Grover's algorithm has been the core quantum ingredient in
the fast quantum decision tree algorithms 
for several other problems. Some examples are,
the algorithm of Boyer et al. \cite{BoyerBHT98}
for computing the $OR$ of $n$ Boolean inputs 
in $O(\sqrt{n})$ queries, the algorithm of Durr and
H{\o}yer \cite{DurrH96} for finding the minimum of
$n$ numbers in $O(\sqrt{n})$ queries, and, more recently,
the algorithm of
Buhrman et al. \cite{BuhrmanDHHMSW00}
for solving the Element Distinctness problem
in $O(n^{3/4} \log{n})$ queries.

The optimality of Grover's algorithm is implied by the
earlier result of Bennett et al. \cite{BennettBBV97}. Beals et al.
\cite{BealsBCMW98} and Ambainis \cite{Ambainis00} prove
some powerful general lower bounds. Notably not all problems
allow quantum speedups. The PARITY function is an example. 

How fast can quantum computers sort? This is probably among the
most natural questions to ask about the power of
quantum computers. It is very natural to study the
comparison-based sorting in the decision tree 
model. This is because the minimum number of comparisons
needed to sort $n$ numbers is just the same as the
decision tree complexity of the following problem: 
given the comparison
matrix of $n$ numbers, output
the order of these numbers. Without loss of
generality, we assume that the comparison operator is
``$\le$''. 

A straightforward information theoretical argument 
gives the $n\log_2{n}$ classical 
lower bound, 
which is matched 
by simple sorting algorithms such as Insertion Sort
with an $O(n)$ additive term. 
Using the quantum algorithm for searching an element in a
sorted list by Farhi et al.\cite{FarhiGGS99} 
as a subroutine in Insertion Sort, a quantum computer needs only
compare $0.526n\log_2{n} + O(n)$ times.
However, the hope of improving this upper bound asymptotically
by improving the upper bound for the ordered search is
made impossible because of
the $\Omega(\log{n})$ lower bound for the latter 
problem due to Ambainis \cite{Ambainis99}. 

Ambainis' lower bound proof for the ordered search
does not seem to apply to the sorting problem.
However, a straightforward application of the same 
author's general lower bound technique \cite{Ambainis00}
gives an $\Omega(n)$ lower bound. 
In this paper,  we improve the lower bound to
$\Omega(n\log{n})$. This means that the best comparison-based
quantum sorting algorithm can be at most a 
constant time faster than the best classical algorithm.
Our approach is similar to that taken by H{\o}yer and Nerbeek 
\cite{HoyerN00} in improving the lower bound for the ordered search
by a constant factor. That is, we use the \emph{weighted} version of
the adversary idea of Ambainis 
\cite{Ambainis00} with a carefully chosen probabilistic distribution
of adversary input pairs, and the lower bound finally relies on 
a property of the Hilbert matrix.

In the next section we give a formal definition of the quantum
decision tree model. After introducing
 some notations, we give a proof overview. 
The complete proof of the main lemma is presented in Section 
\ref{mainlemma}, which is followed by
the section for open problems.

\section{Notations and overview}
\subsection{The model} Following
the paper of Beals et al. \cite{BealsBCMW98}, 
we give a formal definition
of the standard quantum decision tree model for
computing a function 
$f: \bstr{m} \rightarrow \bstr{k}$. 

The quantum algorithm works in the Hilbert space
$$\mathbf{span} \{
      | s, b, c \rangle : \textrm{$ s \in \mathbb{Z}_m$,
				  $b \in \bstr{}$,
				  $c \in \bstr{w}$, for some integer $w \ge 0$}
		      \}.$$
An oracle holds the input 
$x  = x_0 x_1 \cdots x_{m-1} \in \bstr{m}$,
which is not known to the algorithm directly.
Recall that in the classical decision tree model,
the input bit $x_i$ is returned to the algorithm if the index 
$i$ is presented to the oracle. In the quantum setting,
due to the requirement of being unitary for all 
quantum operations, 
$x_i$ is XOR-ed with another bit provided by the algorithm.
Mathematically, the algorithm ``reads'' the input
through the \emph{oracle gate}, which is
a unitary transformation that depends on $x$ and works
on each base vector $|s, b, c\rangle$ as follows:
$$ O_x |s, b, c\rangle = |s, b\oplus{}x_s, c\rangle.$$
It is easy to verify that with 
$\Psi^\pm = \frac{1}{\sqrt{2}}(|0\rangle \pm |1\rangle)$, 
$$O_x |s, \Psi^-, c\rangle = (-1)^{x_s}
          |s, \Psi^-, c\rangle,$$
and,
$$O_x |s, \Psi^+, c\rangle = |s, \Psi^+, c\rangle.$$
A quantum decision tree algorithm that makes $T$
queries is the application of
a sequence of unitary transformations on the initial state 
$|\overrightarrow{0}\rangle$:
$$U_T O_x U_{T-1} O_x \cdots U_1 O_x U_0 |\overrightarrow{0}\rangle.$$
For all $t$, $0\le t\le T$, we write
  $$\state{x}:= U_{t} O_x \cdots U_0 |\overrightarrow{0}\rangle.$$
We say that the quantum algorithm computes $f$
with the error probability bounded by $\epsilon$, for some constant
$\epsilon$ such that $0 \le \epsilon < 1/2$,
if there is a measurement $M$, such that for any input $x$,
when $M$ is applied to $\state[T]{x}$,
the probability of observing $f(x)$ is no less than $1-\epsilon$.

\subsection{More notations} 
For our lower bound purpose, it suffices to assume that
$x_0, x_1, \cdots, x_{n-1}$,
the $n$ numbers to be sorted, 
correspond to
some permutation $\sigma$ of $\{0, 1, \cdots, n-1\}$, i.e.,
$x_i = \sigma(i)$, for all $0\le{}i\le{}n-1$. The input to the quantum
decision tree algorithm is the comparison matrix 
$M_\sigma \in \mathbf{M}_n$, i.e., for all $i$ and $j$, 
$0\le i, j\le n-1$, 
\begin{displaymath}
(M_\sigma)_{i,j} = \left\{
\begin{array}{ll}
1 & \textrm{if $\sigma(i) \le \sigma(j)$,}\\
0 & \textrm{otherwise.}
\end{array}
\right.
\end{displaymath}
    For the simplicity of notations, when the subscript $M_\sigma$
is needed, we use $\sigma$ instead.
    The Hilbert space of the algorithm is now
$$H := \textbf{span}\{
     |i, j, b, c \rangle : \textrm{$0\le i, j \le n-1$,
				    $b \in \bstr{}$,
				    $c \in \bstr{w}$, for some $w\ge 0$}
                          \}.$$
    The oracle gate that corresponds to the input $\sigma$ works as follows:
$$O_\sigma |i, j, b, c\rangle
= |i, j, b{\oplus}(M_\sigma)_{i, j}, c\rangle.$$
     For all $i$ and $j$, $0\le i, j \le n-1$, let $P_{i, j}$ be projection from $H$ to
the subspace
$$\mathbf{span}\{|i, j, \Psi^-, c\rangle, |j, i, \Psi^-, c\rangle\ : 
  \textrm{$c\in{}\bstr{w}$}\}.$$
     For any permutation $\sigma$, and any
$k, d \in \mathbf{Z}$ with
 $0\le k \le n-2$ and $1\le d\le n-k-1$, we define a new 
permutation
$$\sigma^{(k, d)} := (k+d, k+d-1, \cdots, k) \circ \sigma.$$
Note that if $\tau = \sigma^{(k, d)}$,
we have
\begin{displaymath}
\sigma^{-1}(i) = \left\{
  \begin{array}{ll}
    \tau^{-1}(i+d) & i=k,\\
    \tau^{-1}(i-1) & k+1\le i \le k+d,\\
    \tau^{-1}(i)   & \textrm{otherwise.}
  \end{array}
\right.
\end{displaymath}
Also note that $M_\sigma$ and $M_\tau$
differ on only the following pairs of indices
$$ \{\sigma^{-1}(k), \sigma^{-1}(k+i)\}  =
 \{\tau^{-1}(k+d), \tau^{-1}(k+i-1)\},$$
for all $i$ such that $1\le i \le d$.
   The weight function $w(\sigma, \tau)$ is defined
for every pair of permutations:
\begin{displaymath}
w(\sigma, \tau) := \left\{
  \begin{array}{ll}
  1/d & \textrm{if $\tau = \sigma^{(k, d)}$, for some $k$ and $d$,}\\
      & \textrm{such that $0\le k\le n-2$ and $1\le d\le n-k-1,$}\\
  0 & \textrm{otherwise.}
\end{array}
\right.
\end{displaymath}
  Define $H_n := \sum_{i=1}^{n} 1/i$. 

\subsection{Proof overview}
Now we describe the main idea of the adversary technique.
For any pair of the inputs $(x, y)$ such that
$f(x) \ne f(y)$, any algorithm that computes $f$
with high success probability
must separate the two final vectors $\state[T]{x}$
and $\state[T]{y}$ far apart. 
However, it is hard for the algorithm to distinguish
$x$ and $y$ if they are very similar. Therefore,
if there is
a probabilistic distribution of close pairs so that 
on average
the algorithm can only separate the corresponding
vectors by a little amount on each step,
then we can argue that 
the algorithm needs many steps.
The quantity $|\langle \state{x} | \state{y}\rangle|$
is used to measure how close the two vectors are.

The follow quantity is an indication of the algorithm's
progress:
$$s_t := \sum_{\sigma, \tau} {w(\sigma, \tau)}
\left|\langle \state{\sigma} | \state{\tau} \rangle\right|.$$
The idea behind the choice of 
$w(\sigma, \tau)$ is that, for any permutation $\sigma$, 
these permutations $\tau$ obtained from $\sigma$ by 
rotating some $d$ consecutive
elements are close to $\sigma$. 
Moreover, the smaller $d$ is, the harder for the
algorithm to distinguished them. Therefore $w(\sigma, \tau)$
is proportional to $1/d$.

Since the algorithm starts with the same initial state regardless
of the input, by usual calculation we have,
\begin{lemma}\label{innitial}
$s_0 = n! (n H_{n-1} - n + 1)$.
\end{lemma}

It is a well known fact 
that if an algorithm computes a function $f$ with the
error probability bounded
by $\epsilon$, then for any pair of 
inputs $x$ and $y$ such that $f(x) \ne f(y)$,  
$$|\langle \state[T]{x} | \state[T]{y} \rangle| \le 2\sqrt{\epsilon(1-\epsilon)}.$$
See, for example, the paper of Ambainis \cite{Ambainis00} for a proof.  
Henceforth,
\begin{lemma}\label{final}
$s_T \le 2\sqrt{\epsilon(1-\epsilon)} s_0$.
\end{lemma}

Our main lemma says that any algorithm can make only a
little bit of progress
in decreasing $s_t$ on each step:
\begin{lemma}[Main Lemma]\label{change}
$|s_{t+1} - s_{t}| \le 2\pi n!$, for all $0\le t \le T-1$.
\end{lemma}

Since 
$$(1-2\sqrt{\epsilon(1-\epsilon)})s_0 \le |s_T - s_0| \le 
	\sum_{t=0}^{T-1}|s_{t+1} - s_0|,$$
by the previous lemmas, we obtain our main theorem:
\begin{theorem}[Main Theorem]
Any quantum sorting algorithm with the
error probability bounded by 
$\epsilon$ must compare no less than
$$\frac{1 - \sqrt{2\epsilon(1-\epsilon)}}{2\pi} (nH_{n-1} - n +1) =
\Omega(n\log{n})$$ times.
\end{theorem}
Setting $\epsilon = 0$, we obtain 
\begin{corollary}
Any error-less quantum sorting algorithm must compare at least
$$\frac{1}{2\pi}n\ln{n} - \frac{1-C_E}{2\pi}n  + O(1)
\approx 0.110 n\log_2{n} - 0.067n + O(1)$$
times, where $C_E = 0.57721566\cdots$ is the
Euler-Mascheroni Constant.
\end{corollary}

Let $A = [\alpha_{k, l}]_{1\le k, l<\infty}$
be the Hilbert matrix with
$\alpha_{k, l} = 1/(k+l-1)$, and
$\snorm{\cdot}$  be the spectral norm, i.e.,
for any complex matrix $M \in \mathbf{M}_m$,
$$\snorm{M} := \max_{x\in\mathbb{C}^m, \|x\|_2=1} \|Mx\|_2.$$
Our proof for the Main Lemma relies on the following
property of the Hilbert matrix:
\begin{lemma} \label{choi}$\snorm{A} = \pi.$
\end{lemma}
Choi \cite{Choi83} has an elegant proof for this lemma.

\section{Proof of the Main Lemma}
\label{mainlemma}
\begin{proof}[Lemma~\ref{change}]
For each individual pair $(\sigma, \tau)$, where
$\tau = \sigma^{(k, d)}$, by the definitions of
$\state[t+1]{\sigma}$ and $\state[t+1]{\tau}$, we have
\begin{eqnarray*}
&&\left| \langle \state[t+1]{\sigma} | \state[t+1]{\tau}\rangle
           -\langle \state{\sigma} | \state{\tau}\rangle 
     \right|\\
&=& \left| \left\langle U_{t+1} 
               (O_\sigma\state{\sigma} 
                    - O_\tau\state{\sigma}
               ) +
              U_{t+1} O_\tau\state{\sigma}
            \left| U_{t+1} O_\tau\state{\tau}\right.\right\rangle
           -\langle \state{\sigma} | \state{\tau}\rangle 
     \right|.
\end{eqnarray*}
By the properties of the inner product and that $U_{t+1}$
is unitary, the above expressions are simplified to
\begin{equation}
 \left| \langle  
               (O_\tau O_\sigma - I)\state{\sigma} 
            | \state{\tau}\rangle
     \right|.
\label{diff} 
\end{equation}
The effects of
$O_\sigma$ and $O_\tau$ cancel out on most base vectors, except  
for these $|i, j, \Psi^{-}, c\rangle$ such that
$(M_\sigma)_{i, j} \ne (M_\tau)_{i, j}$. Therefore
Eq.~\ref{diff} is bounded from the above by
$$ 2\sum_{i=1}^d 
    \left| \langle
           P_{\pos{k}, \pos{k+i}} 
           \state{\sigma} | \state{\tau}
           \rangle
    \right|.$$
By the Cauchy-Schwarz inequality, this expression can be
further upper-bounded by
\begin{equation}
 2 \sum_{i=1}^{d} 
     \left\|P_{\pos{k}, \pos{k+i}} \state{\sigma} \right\| 
          \cdot
     \left\|P_{\pos{k}, \pos{k+i}} \state{\tau}\right\|.
\label{eachupp}
\end{equation}

Now we are ready to bound $\Delta_t := |s_{t+1} - s_t|$. By definitions
and pulling summations out of the absolute value, we obtain
\begin{equation}
\Delta_t \le
  \sum_{\sigma} \sum_{k=0}^{n-2} \sum_{d=1}^{n-k-1}\frac{1}{d}
     \left| \langle \state[t+1]{\sigma} | \state[t+1]{\sigma^{(k, d)}}\rangle
           -\langle \state{\sigma} | \state{\sigma^{(k, d)}}\rangle
     \right|.
\end{equation} 
Plug in the upper bound of Eq.~\ref{eachupp}, 
this is then upper-bounded by
$$ 2 \sum_{\sigma} \sum_{k=0}^{n-2} \sum_{d=1}^{n-k-1}
  \sum_{i=1}^d \frac{1}{d}
	\left\|P_{\pos{k}, \pos{k+i}} \state{\sigma} \right\| \cdot
               \left\|P_{\pos{k}, \pos{k+i}} \state{\sigma^{(k, d)}}\right\|.$$
By reordering the terms of the summation and applying the
Cauchy-Schwarz inequality, this
is further upper-bounded by
\begin{equation}
 2 \sum_{d=1}^{n-1} \frac{1}{d} \sum_{i=1}^{d}
     \sqrt{\sum_{\sigma}\sum_{k=0}^{n-d-1}
      \left\|P_{\pos{k}, \pos{k+i}} \state{\sigma} \right\|^2}
      \sqrt{\sum_{\sigma}\sum_{k=0}^{n-d-1}
      \left\|P_{\pos{k}, \pos{k+i}} \state{\sigma^{(k, d)}}\right\|^2}.
\label{bound1}
\end{equation} 
Let $a = [a_i]_{1\le i \le n-1} \in \mathbb{C}^{n-1}$
be a column vector with
$$a_i := \sqrt{\sum_\sigma
     \sum_{l=0}^{n-i-1} \left\|P_{\sigma^{-1}(l), \sigma^{-1}(l+i)} \state{\sigma}\right\|^2}.$$
Clearly,
  $$\sqrt{\sum_{\sigma}\sum_{k=0}^{n-d-1}
      \left\|P_{\pos{k}, \pos{k+i}} \state{\sigma} \right\|^2}
      \le a_i,$$
and,
\begin{eqnarray*}
&&  \sqrt{\sum_{\sigma}\sum_{k=0}^{n-d-1}
      \left\|P_{\pos{k}, \pos{k+i}} \state{\sigma^{(k, d)}}\right\|^2}\\
&=&       \sqrt{\sum_{\tau}\sum_{k=0}^{n-d-1}
      \left\|P_{\pos[\tau]{k+d}, \pos[\tau]{k+i-1}} 
	\state{\tau}\right\|^2}\\
&\le& a_{d-i+1}.
\end{eqnarray*}

Let $K_{n-1} = [\kappa_{k, l}]_{1\le k, l\le n-1} \in \mathbf{M}_{n-1}$ 
be a Hankel matrix with 
\begin{displaymath}
\kappa_{k, l} = \left\{
   \begin{array}{ll}
  1/{(k+l-1)} & \textrm{$k+l \le n$,}\\
  0 & \textrm{otherwise.}
   \end{array}
  \right.
\end{displaymath}
Now Eq.~\ref{bound1} can be upper-bounded by
\begin{equation}
  2 \sum_{d=1}^{n-1}\sum_{i=1}^{d}\frac{1}{d} a_i a_{d-i+1}
= 2 a^T K_{n-1} a.
\end{equation}
Since every $\state{\sigma}$ is a unit vector, we have
\begin{eqnarray*}
\|a\|_2^2 &=& \sum_{i=1}^{n-1} \sum_\sigma \sum_{l=0}^{n-i-1}
              \|P_{\sigma^{-1}(l), \sigma^{-1}(l+i)} \state{\sigma}\|^2\\
&=& \sum_\sigma \sum_{l=0}^{n-2} \sum_{i=1}^{n-l-1}
              \|P_{\sigma^{-1}(l), \sigma^{-1}(l+i)} \state{\sigma}\|^2\\
&\le&  n!.
\end{eqnarray*}
Clearly $\snorm{K_{n-1}} \le \snorm{A} = \pi$,
by Lemma~\ref{choi}. 
Applying the Cauchy-Schwarz inequality and 
by the definition of the spectral norm,  we obtain the desired
upper bound:
$$|s_{t+1} - s_{t}| 
 \le  2 \|a\|^2_2 \snorm{K_{n-1}}
 \le  2 \pi n!.$$
\end{proof}

\section{Open problems}
The result of Grigoriev et al. \cite{GrigorievKHS96}
implies that if only comparisons are allowed, 
the randomized decision tree complexity of 
Element Distinctness has the same 
$\Omega(n\log{n})$ lower bound as the 
sorting problem. 
Interestingly, their quantum complexities differ 
dramatically: the quantum algorithm for Element 
Distinctness due to Buhrman et al. \cite{BuhrmanDHHMSW00}
compares only
$O(n^{3/4}\log{n})$ times. It would be interesting
to improve the current trivial lower bound of
$\Omega(\sqrt{n})$ for Element Distinctness.

The decision tree complexity of nontrivial 
monotone graph properties has been a
classical subject. Recently, Yao \cite{Yao00}
proves an $\Omega(n^{2/3})$ lower bound
for the quantum complexity of every
nontrivial monotone graph property.
It is reasonable to conjecture that the correct
general quantum lower bound should be $\Omega(n)$.
The best known lower bounds for specific 
properties, such as the $\Omega(n)$ lower bound for
Connectivity implied by Ambainis' \cite{Ambainis00} 
general technique,
do not seem to be tight. It
would be interesting to improve both the
general and the specific lower bounds. Probably this will
require new techniques other than that of
Ambainis \cite{Ambainis00}.

Space-time tradeoffs for sorting
and related problems have been studied 
for the classical case. A $Time \cdot Space$
lower bound of $\Omega(n^2)$ is proved for the
comparison-based
sorting by Borodin et al. \cite{BorodinFKLT81},
and for the R-way branching program by
Beame \cite{Beame91}. Formulations and results on
the quantum time-space tradeoffs for sorting and other
problems such as Element Distinctness would be interesting. 

\section{Acknowledgment}
The author would like to thank Andy~Yao for discussions,
Dieter~van~Melkebeek, Daniel~Wang, and 
especially Sanjeev~Arora,
for their precious comments and suggestions.


\begin{thebibliography}{10}

\bibitem{STOC28}
{\em Proceedings of the Twenty-Eighth Annual ACM Symposium on the Theory of
  Computing}, Philadelphia, Pennsylvania, May 1996.

\bibitem{Ambainis99}
A.~Ambainis.
\newblock A better lower bound for quantum algorithms searching an ordered
  list.
\newblock In {\em 40th Symposium on Foundations of Computer Science ({FOCS})},
  pages 352--357, New York, NY, USA, October 1999. IEEE Computer Society.

\bibitem{Ambainis00}
A.~Ambainis.
\newblock Quantum lower bounds by quantum arguments.
\newblock In {\em Proceedings of the Thirty-second Annual ACM Symposium on the
  Theory of Computing}, pages 636--643, Portland, Oregon, May 2000.

\bibitem{BealsBCMW98}
R.~Beals, H.~Buhrman, R.~Cleve, M.~Mosca, and R.~de~Wolf.
\newblock Quantum lower bounds by polynomials.
\newblock In {\em 39th Annual Symposium on Foundations of Computer Science},
  pages 352--361, Los Alamitos, CA, November 1998. IEEE.

\bibitem{Beame91}
Paul Beame.
\newblock A general sequential time-space tradeoff for finding unique elements.
\newblock {\em SIAM J. Comput.}, 20(2):270--277, 1991.

\bibitem{BennettBBV97}
Charles~H. Bennett, Ethan Bernstein, Gilles Brassard, and Umesh Vazirani.
\newblock Strengths and weaknesses of quantum computing.
\newblock {\em SIAM Journal on Computing}, 26(5):1510--1523, October 1997.

\bibitem{BorodinFKLT81}
Allan Borodin, Michael~J. Fischer, David~G. Kirkpatrick, Nancy~A. Lynch, and
  Martin Tompa.
\newblock A time-space tradeoff for sorting on nonoblivious machines.
\newblock {\em J. Comput. System Sci.}, 22(3):351--364, 1981.
\newblock Special issue dedicated to Michael Machtey.

\bibitem{BoyerBHT98}
M.~Boyer, G.~Brassard, P.~H{\o}yer, and A.~Tap.
\newblock Tight bounds on quantum searching.
\newblock {\em Fortschritte der Physik}, 46(4-5), 1998.

\bibitem{BuhrmanDHHMSW00}
Harry Buhrman, Christoph Durr, Mark Heiligman, Peter H{\o}yer, Frederic
  Magniez, Miklos Santha, and Ronald de~Wolf.
\newblock Quantum algorithm for element distinctness.
\newblock Preliminary version: quant-ph/0007016, 2000.

\bibitem{Choi83}
M.-D. Choi.
\newblock Tricks or treats with the {H}ilbert matrix.
\newblock {\em Amer. Math. Monthly}, 90(5):301--312, 1983.

\bibitem{DurrH96}
Christoph Durr and Perter H{\o}yer.
\newblock A quantum algorithm for finding the minimum.
\newblock Preliminary version: quant-ph/9607014, 1996.

\bibitem{FarhiGGS99}
Edward Farhi, Jeffrey Goldstone, Sam Gutmann, and Michael Sipser.
\newblock Invariant quantum algorithms for insertion into an ordered list.
\newblock Preliminary version: quant-ph/9901059, 1999.

\bibitem{GrigorievKHS96}
Dima Grigoriev, Marek Karpinski, Friedhelm Meyer auf~der Heide, and Roman
  Smolensky.
\newblock A lower bound for randomized algebraic decision trees.
\newblock In ACM \cite{STOC28}, pages 612--619.

\bibitem{Grover96}
Lov~K. Grover.
\newblock A fast quantum mechanical algorithm for database search.
\newblock In ACM \cite{STOC28}, pages 212--219.

\bibitem{HoyerN00}
P.~H{\o}yer and J.~Neerbek.
\newblock Bounds on quantum ordered searching.
\newblock Preliminary version: quant-ph/0009032, 2000.

\bibitem{Shor97}
Peter~W. Shor.
\newblock Polynomial-time algorithms for prime factorization and discrete
  logarithms on a quantum computer.
\newblock {\em SIAM Journal on Computing}, 26(5):1484--1509, October 1997.

\bibitem{Yao00}
Andrew Yao.
\newblock Personal communication, 2000.

\end{thebibliography}
\end{document}